\documentstyle[floats,aps,aipbook,psfig]{revtex}
\righthead{Norris {\it et al.}}
\lefthead{Time Dilation in Pulse Structures}
\begin{document}
\title{Calibration of Tests for Time Dilation in GRB Pulse Structures}
\author{J.P. Norris,$^1$ R.J. Nemiroff,$^2$ 
   J.T. Bonnell,$^3$ and J.D. Scargle$^4$}
\address{$^1$NASA/Goddard Space Flight Center, Greenbelt, MD 20771\\
         $^2$George Mason University, Fairfax, VA 22030\\
         $^3$Universities Space Research Association\\
         $^4$NASA/Ames Research Center, Moffett Field, CA 94035}

\maketitle

\begin{abstract}
Two tests for cosmological time dilation in $\gamma$-ray bursts -- the peak 
alignment and auto-correlation statistics -- involve averaging information 
near the times of peak intensity.  Both tests require width corrections, 
assuming cosmological origin for bursts, since narrower temporal structure 
from higher energy would be redshifted into the band of observation, and since 
intervals between pulse structures are included in the averaging procedures.  
We analyze long ($>$ 2 s) BATSE bursts and estimate total width corrections for 
trial time-dilation factors (TDF = [1+$z_{\rm dim}$]/[1+$z_{\rm brt}$]) by 
time-dilating and redshifting bright bursts.  Both tests reveal significant 
trends of increasing TDF with decreasing peak flux, but neither provides 
sufficient discriminatory power to distinguish between actual TDFs in the 
range 2--3.
\end{abstract}

\section*{Two Width Corrections to Observed Time Dilation}
If $\gamma$-ray bursts (GRB) are at cosmological distances, then the temporal 
structure associated with each energy range at the source is shifted to lower 
energies in the observer's frame of reference.  Since GRB pulse structures are 
narrower at higher energy, this redshift-dependent {\it narrowing} would 
compete with cosmological time dilation.  Note that both redshift-dependent 
narrowing and time dilation have mathematical analogues in special relativistic 
(SR) beaming models -- blue-shifting of the radiation, and time contraction of 
the temporal structure.  So far, there is no observational data which affords a 
way to distinguish between SR and cosmology in GRBs.  We discuss the problem in
terms of the cosmological hypothesis, keeping in mind that both cases have the
same temporal mensuration problems.

Each measure of time dilation in GRBs requires a different width correction for
narrowing of temporal structure, but all such corrections operate in the same
sense:  The {\it actual} time-dilation factor (TDF) is decreased by the 
redshift effect, so that the {\it observed} TDF, between bright and dim groups 
of bursts, is smaller, and a function of temporal structure in both groups,

$$ TDF_{\rm obs} = F[ \ TDF_{\rm act}, \ 
           \Lambda ( TDF_{\rm act} \ \times \ E_c ), \ \Lambda (E_c) \ ]
\eqno(1)$$
where $E_{\rm c}$ is the energy band of observation, and $TDF_{\rm act}$ 
$\times$ $E_{\rm c}$ is the band, relative to $z_{\rm brt}$ of bright bursts, 
from whence the temporal structure was redshifted \cite{{JPN1},{JPN2},{FandB}}.
Of course, $TDF_{\rm act}$ = [1+$z_{\rm dim}$]/[1+$z_{\rm brt}$]) 
is what we are after.  The insidious part is the dependence of $\Lambda$, a 
generic width statistic, on $TDF_{\rm act}$.  Thus the energy-dependent 
width correction for redshift is like a ratio of widths of 
temporal structure, but it is not a simple ratio for all time dilation 
measures, for the following reason:  For measures which utilize information 
from a portion of the time profile that contains any intervals or valleys 
between pulse structures, time dilation of these regions interjacent to peaks 
``subtracts'' from time dilation of regions of emission, since interval regions 
and emission regions are not segregated.  Thus when these portions of time 
profiles are averaged -- as in Peak Aligned profile (PA) and Auto-Correlation 
Function (ACF) measures -- regions of emission and intervals between pulses 
are thrown together in the averaging process.  The resulting situation is 
illustrated in Figure 4 of ref \cite{JPN3}, in these proceedings.

For the PA and ACF statistics we have calibrated the resulting diminution of
time dilation that would arise for the combined effects of redshift of
temporal structure and averaging of regions that contain intervals.  
Note that correction for the latter effect, previously not
addressed \cite{{JPN1},{FandB}}, is not required for time-dilation measures 
which rely upon distributions of a measured parameter, such as distributions of 
pulse widths, intervals between pulses, or burst durations.  Instead, these 
measures have simple width-correction ratios, that can be obtained from
the distributions in two relevant energy bands.

\section*{Calibration of Width Corrections}

The PA and ACF tests for time dilation have been described before 
\cite{{JPN1},{FandB},{Mitro}}.  The basic procedures are:  divide bursts into 
groups based on some measure of brightness; and within a brightness group,
average the profiles with their highest intensity peaks in registration, or
average the profiles' ACFs.  The ``common-sense'' appeals of the PA test are 
that it operates entirely in the time domain, and makes use of the most intense 
part of a burst.  The efficacy of the ACF test is that it probes short 
timescales, to the limit of temporal resolution, without the need to worry 
about finding the exact location of peak intensity in dim bursts.  Also, 
the ACF has a well-defined correction for co-added noise at zero lag.  For 
both tests the main problem is that the width corrections are appreciable, 
resulting in less than satisfactory discriminatory power in the $TDF_{\rm act}$ 
range $\sim$ 2--3, given the present sample variance.

\paragraph*{Data preparation.}   
We use BATSE DISCSC data summed over channels 1 and 2 ($\sim$ 27--115 keV) to 
construct average peak-aligned profiles and ACFs for bursts longer than 
$\sim$ 2 s and with peak intensities higher than 1400 counts s$^{-1}$.  
The bursts are divided into six brightness groups, $\sim$ 85 bursts 
per group, according to their BATSE 3B peak fluxes determined at 256 ms.  This
timescale compromises between 64 ms, where noisier estimates are obtained for
dim bursts, and 1024 ms, which integrates over pulse widths (pulses in long
bursts having FWHM $\sim$ 100--500 ms, dependent on energy band \cite{JPN4}).  
Quadratic (infrequently, higher order) backgrounds are fitted and subtracted.  
For the PA test the time profiles are rendered to 512-ms resolution; the 
original 64-ms resolution is preserved for the ACF test.  To approximately 
nullify brightness bias, the comparison of average profiles or ACFs between 
different brightness groups (described below) is performed with the 
signal-to-noise (s/n) levels of the individual time profiles of the bright 
group equalized to the s/n levels of the profiles of the other groups.  For 
each of the five brightness groups below the brightest, ten such noisy 
realizations per bright burst are computed, with the new peak intensity chosen 
randomly from among the peak intensities of the bursts of each respective 
group.  Thus, about 850 $\times$ 5 = 4250 noisy realizations of bright bursts 
are created.

\paragraph*{Estimating observed time dilation.}  
For -8 to +16 (512-ms) bins of the peak of the average PA profiles of the
bright bursts, the intensity levels and corresponding profile widths for a 
dimmer group are found, and width ratios computed for the 24 bins 
($N_{\rm PA}$).  A similar procedure is followed for the ACFs, for $\pm$60 
64-ms lag bins ($N_{\rm ACF}$).  However, since pulse widths (FWHM) in 
individual bursts are $\sim$ 500 ms, only $N_{\rm indep}$ $\approx$ 12 and 4 
independent ratio estimates result for PA and ACF procedures, respectively.
We estimate means, standard and sample errors by a bootstrap procedure 
\cite{Efron}.  The $\sim$ 85 burst profiles (or ACFs) in each brightness group 
(850 noisy realizations for the brightest group) are considered the ``parent 
population'' from which 85 profiles are drawn randomly with replacement.  For 
each brightness group the random selection is repeated 500 times.  For each run 
the average profile is computed, and the width ratios computed as described 
above.  The 500 $\times$ $N_{\rm PA}$ (or $N_{\rm ACF}$) width ratios are rank 
ordered.  The resulting 50$^{\rm th}$, 15$^{\rm nd}$, and 84$^{\rm th}$ 
percentile levels are taken as the median width ratio -- box symbols in 
Figures 1 and 2 -- negative and positive 1-$\sigma$ standard errors, 
respectively.  The sample errors plotted are these 1-$\sigma$ errors reduced 
by the factor $\sqrt {N_{\rm indep}}$.

For the PA and ACF measures, the {\it observed} TDFs range up to $\sim$ 1.75 
and $\sim$ 1.45, respectively, for the dimmest group relative to the brightest. 
But the {\it actual} TDFs would be larger in both the cosmological and SR 
beaming hypotheses, by width-correction factors we estimate as follows.

\paragraph*{Estimating width corrections, interval dilation+redshift effects.}
The expected time-dilation ``signal'', {\it sans} redshift effect, is easily 
simulated by merely stretching the profiles (using the original 64-ms data) of 
the bright group by factors of 2.0 and 3.0, and comparing with the unstretched
profiles, but now using 16-channel MER data.  For {\it actual} TDFs of 2.0 and 
3.0, the {\it recovered} TDFs for the PA measure are $\sim$ 1.7 (85\%) and 
$\sim$ 2.55 (85\%), respectively (circle symbols, Figure 1).  Similarly for 
the ACF measure, the recovered TDFs are $\sim$ 1.6 (80\%) and $\sim$ 2.0 
(66\%), respectively (circle symbols, Figure 2).  Incomplete recovery of the 
input TDF is attributable to inclusion of regions which contain stretched 
intervals as well as stretched pulse structures.

\begin{figure}
\leavevmode
\psfig{file=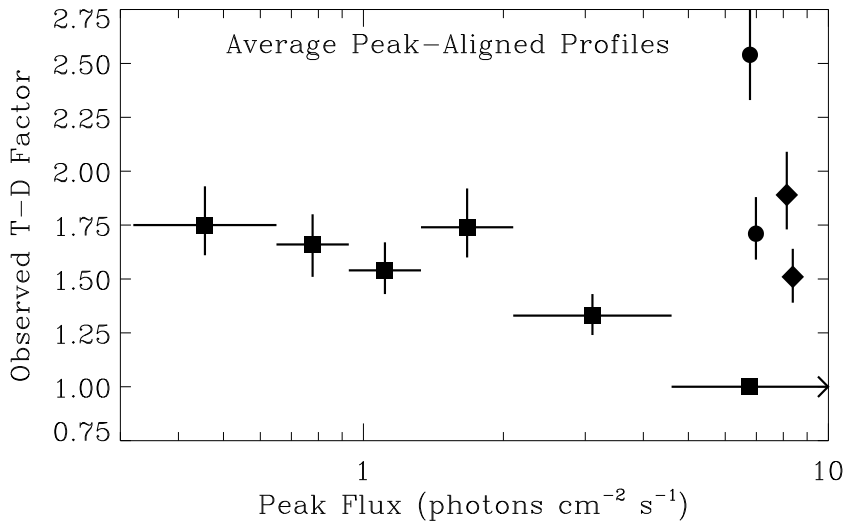,height=3.0in,width=5.0in}
\caption{
Box symbols indicate observed time-dilation factor vs. BATSE 3B peak
flux (256 ms), determined from average peak-aligned profiles.  Errors 
estimated by bootstrap method.  On right side, upper four points (without 
horizontal bars) are the bright burst sample time-dilated (circle symbols) and 
then redshifted as well (diamond symbols), by factors of 2 and 3 (lower and 
upper symbols, respectively).  Approximately 85 bursts per group.}
\end{figure}
Recall that we analyzed the six brightness groups in the 25--115 keV band (box
symbols, Figures 1 and 2).  The additional effect of redshift of temporal 
structure is simulated by using the 16-channel data for bright bursts:  
corresponding approximately to TDF=1, $\Sigma$ chans: 2--6, $\sim$ 22--100 keV; 
TDF=2, $\Sigma$ chans: 4--8 + half of 9, $\sim$ 41--200 keV; and TDF=3, 
$\Sigma$ chans: half of 5 + 6--10, $\sim$ 65--315 keV.  The combined effect of 
stretching and redshifting profiles of bright bursts is indicated by 
diamond symbols in Figures 1 and 2.  Narrower structure redshifted into the 
band of observation further reduces the observed TDF.  For actual TDFs and 
redshifts of 2.0 and 3.0, the recovered TDFs are now $\sim$ 1.5 (75\%) and 
$\sim$ 1.9 (63\%), respectively, for the PA measure; the corresponding values 
for the ACF measure are $\sim$ 1.4 (70\%) and $\sim$ 1.6 (54\%), respectively.  
As can be seen by comparing the pairs of diamond symbols (Figures 1 and 2) 
with the observed TDF determinations for the six brightness groups, 
the uncertainties are such that $TDF_{\rm act}$ is only constrained to
the range $\sim$ 2--3 for the dimmer groups via both the PA and ACF measures.

In conclusion, on the short (64 ms -- few s) and intermediate (1--20 s) 
timescales probed by the PA and ACF tests, observed TDFs, relative to 
bright bursts, range up to $\sim$ 1.45 (ACF) and $\sim$ 1.75 (PA).  
From calibrations using the bright sample we conclude that, for the
same input time-dilation and redshift factor, the ACF {\it is expected} to 
yield smaller $TDF_{\rm obs}$.  In fact, actual cosmological TDFs would be 
somewhat larger than $TDF_{\rm obs}$:  Two effects, redshift of narrower 
structure into the band of observation, and inclusion of stretched intervals, 
result in smaller observed time-dilation factors.  The second effect was not 
appropriately simulated in previous estimates which used ratios of average 
pulses \cite{JPN1} or ratios of average ACFs \cite{FandB} in different energy 
bands of bright bursts.  The width corrections are more pronounced at higher 
TDFs, such that with present uncertainties, both the PA and ACF measures only 
constrain $TDF_{\rm act}$ = [1+$z_{\rm dim}$]/[1+$z_{\rm brt}$] to lie in the 
range 2--3.

\begin{figure}
\leavevmode
\psfig{file=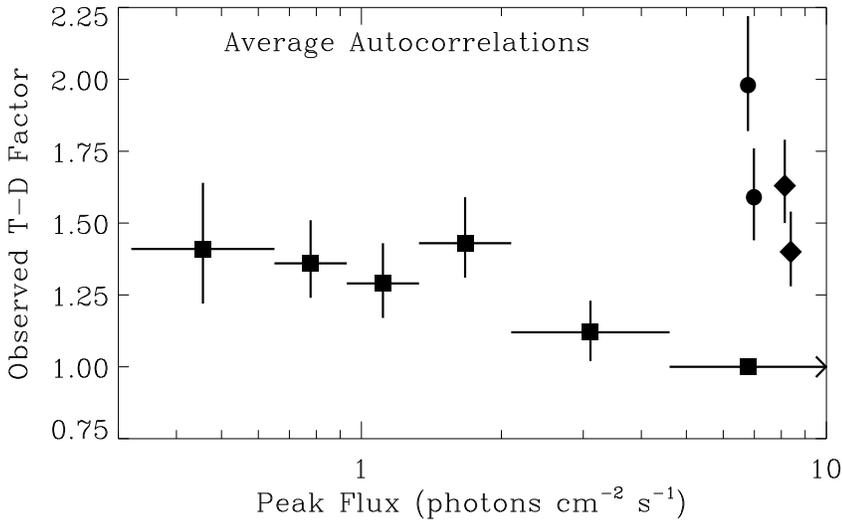,height=3.0in,width=5.0in}
\caption{
Similar to Figure 1, except determined from the ACFs of bursts in six 
brightness groups.  Observed TDFs (boxes) are lower than for peak-aligned 
profiles, but calibrations obtained by time-dilating (circles) and then 
redshifting (diamonds) bright bursts by factors of 2 and 3 are lower as well.
Relatively larger error bars result than for PA measure since fewer independent 
time bins were used.}
\end{figure}

\end{document}